\documentclass[letter]{aa}  
\usepackage{threeparttable} 
\usepackage{graphicx}
\usepackage{gensymb}
\usepackage{amsmath}
\usepackage{txfonts}
\begin{document}

   \title{Ultra-fast simulations of the solar dipole and open flux}
   
   \author{Ismo Tähtinen
                 \inst{1}
          \and
          Timo Asikainen\inst{1}
          \and
          Kalevi Mursula\inst{1}
          }
   \institute{Space Physics and Astronomy Research Unit, University of Oulu,
               POB 8000, FI-90014, Oulu, Finland\\
              \email{ismo.tahtinen@oulu.fi}
             }


 
  \abstract
   {Solar dipole captures important information about the large-scale solar magnetic field.
   The evolution of the solar magnetic field including the solar dipole can be simulated with a surface flux transport (SFT) model, but these simulations are more extensive than is necessary to produce the evolution of the dipole alone.
  }
     {We present a dipole flux transport (DFT), matrix method that combines the classic SFT model with dipole vector representation of the solar magnetic field, allowing significantly faster simulations of the solar dipole.
  }
     {By simulating the evolution of basis vectors of a synoptic map, we constructed propagator matrices that produce the time evolution of the solar magnetic field by means of matrix multiplication.
     The computational speedup is achieved by compressing the propagator matrices to very small fraction $(< 10^{-4}$) of their original size with a recent vector sum method.
  }
     {Depending on time resolution, the DFT performs 100-1000 times faster than a 4-year SFT simulation of a single active region while producing equivalent results.
     For multiple source regions, daily propagation matrices are sufficient to produce results that agree within 1\% with the SFT simulation of solar cycle 24, while performing 80 times faster.
     If the evolution of individual active regions is needed, the DFT performs 50000 times faster than the SFT model.
  }
     {DFT makes solar dipole simulations extremely fast, making it possible to run thousands of simulations in a few minutes with a basic laptop setup.
     As the magnitude of the dipole vector closely matches with open solar flux (OSF) from the potential field source surface model, the DFT can be used to study the development of OSF in various scenarios extremely efficiently.
  }


   \maketitle
%

\section{Introduction}\label{sec:Intro}
The evolution of the solar magnetic field is routinely simulated using surface flux transport (SFT) models \citep[for a recent review and references, see][]{Yeates2023}.
For some applications it is not necessary to retain all information about the simulated photospheric magnetic field, but it suffices to approximate the magnetic field as a global dipole.
In particular, the axial dipole moment of the Sun has been studied extensively, because its strength around the solar minimum correlates quite well with the strength of the following solar cycle \citep{Upton2014,Cameron2016,Ijima2017,Jiang2018,Whitbread2018,Yeates2020,Petrovay2020}.
More generally, the solar dipole is also relevant to studies of coronal and heliospheric magnetic fields, because the higher order multipoles fall off rapidly with height \citep{Wang2009}.
For example, the equatorial dipole moment has been shown to play an important role in shaping the heliospheric magnetic field \citep{WangSheeley2000a,WangSheeley2000b,WangSheeley2003}, which in turn modulates the propagation of cosmic rays in the heliosphere \citep{Jokipii1981}.

In certain situations, SFT calculations can be simplified or even avoided entirely by considering the symmetries of the problem.
When the quantity of interest is the axial dipole moment, the full 2D-simulation can be reduced to a 1D-simulation since the axial dipole depends only on the longitudinally averaged field \citep{DeVore1984,Cameron2007,Ijima2017,Petrovay2019,Yeates2020}.
Furthermore, if only the asymptotic behavior of the axial dipole moment is required, SFT calculations can be replaced by an algebraic approach \citep{Petrovay2020,Wang2021}.
This method maps the magnetic field of an active region to its asymptotic axial dipole moment, exploiting the fact that in SFT simulations the field approaches a steady state in which the magnetic flux is confined to the polar regions.
The asymptotic axial dipole moment of an active region can be expressed as the initial axial dipole strength multiplied by an amplification factor.
For bipolar magnetic regions (BMRs), this factor is a Gaussian function of latitude alone, but the method can be generalized to arbitrary flux distributions \citep{Jiang2014,Petrovay2020,Wang2021,Talafa2022,Yeates2023}.
Furthermore, because the SFT model is linear, individual contributions can be summed to obtain the total axial dipole moment.
In contrast, \citet{Tahtinen2026}, who studied the evolution of the equatorial rather than the axial dipole, avoided performing $>10^5$ SFT simulations by rotating the solutions of previously simulated active regions in longitude.
Due to the linearity and longitudinal symmetry of the SFT model, these rotated solutions are equivalent to simulations with correspondingly rotated source regions.

The strength of the solar dipole is typically modeled with solar dipole moment corresponding to $l=1$ term in spherical harmonic expansion.
Recently, \citet{Tahtinen2024,Tahtinen2026} developed an alternative way for describing the solar dipole.
Their vector sum method produces a vector which shares orientation with the traditional dipole, but whose magnitude has units of magnetic flux and equals the total photospheric magnetic flux aligned with the dipole axis.
Importantly, the magnitude of this dipole vector closely matches with the open solar flux (OSF) from the potential field source surface (PFSS) model  \citep{Altschuler1969,Schatten1969,Wang1992} with standard source surface radius of $2.5R_\odot$.
This result directly relates the total magnetic flux escaping from the Sun (i.e., OSF) to the photospheric magnetic flux distribution, at least in the context of PFSS modeling.

In this Letter, we show how the vector sum method of \citet{Tahtinen2024,Tahtinen2026} can be combined with a classic SFT model to construct a propagator matrix that maps the initial magnetic field of a synoptic map to a dipole vector at a later time.
We refer to this matrix method as dipole flux transport (DFT) as it describes the evolution of the solar dipole vector under the dynamics of the SFT model.
DFT offers a speedup of up to 4 orders of magnitude in our test cases representing various problems in solar dipole modeling.
As the magnitude of the dipole vector closely matches with the OSF from the PFSS model, the method also provides a way to study the development of OSF extremely efficiently.

\section{Data}\label{sec:Data}
We compare the performance of the DFT to the SFT model by simulating the evolution of solar cycle 24.
We use the pole-filled synoptic maps of the radial magnetic field (\texttt{hmi.synoptic\_mr\_polfil\_720s}) from the Helioseismic and Magnetic Imager on board the Solar Dynamics Observatory \citep[SDO/HMI;][]{Scherrer2012,Pesnell2012} and the active region database derived from Spaceweather HMI Active Region Patch data \citep[SHARP;][]{Bobra2014} using an open-source Python code\footnote{\url{https://github.com/antyeates1983/sharps-bmrs}} described in \citet{Yeates2020}.
This algorithm transforms HMI SHARPS into uniform sine-latitude and longitude grid and removes active regions with large flux imbalance.
Active regions that are recognized as remnants of regions observed in previous rotation are also removed.
Data covers Carrington rotations 2097--2224 (May 2010 to December 2019) that correspond to solar cycle 24.
Altogether there are 1095 active regions.

\section{Dipole flux transport}\label{sec:Method}
Our aim is to construct a propagator matrix that represents the evolution of the solar dipole vector under the dynamics of the SFT model described in Appendix~\ref{appendix:SFT}.
The idea is to first construct the propagator \textbf{M} for the full SFT simulation, which turns the SFT simulation into matrix multiplication 
\begin{equation}\label{eq:SFTProduct}
\mathbf{B}(t) = \textbf{M}(t)\mathbf{B_0},
\end{equation}
where the propagator \textbf{M}($t$) maps the initial magnetic field $\mathbf{B_0}$ to final magnetic field \textbf{B} at time $t$.
The matrix \textbf{M} is inconveniently large for offering a significant speedup for full SFT simulations, but its size can be reduced by four orders of magnitude using the vector sum method described in Appendix~\ref{appendix:VectorSum}.
This compactified matrix can then be used to simulate the time evolution of the solar dipole as
\begin{equation}\label{eq:VSProduct}
\mathbf{D}(t) = \textbf{VM}(t)\mathbf{B_0},
\end{equation}
where \textbf{V} represents the vector sum matrix that maps the full synoptic map into a dipole vector \textbf{D}.
Matrix \textbf{VM} then represents the propagator that maps the initial magnetic field $\mathbf{B_0}$ to dipole vector $\textbf{D}$ at some later time $t$.
Next we show in detail how to construct the propagator matrix \textbf{VM}.

\subsection{Propagator for the surface flux transport}
Let $E_{ij}$ denote a matrix unit which has only one nonzero value with value 1 at $i$th row and $j$th column.
These matrix units $E_{ij}$ correspond to standard basis vectors of synoptic map.
A full synoptic map can then be written as
\begin{equation}
B = \sum_{ij} B^{ij}E_{ij},    
\end{equation}
where $B^{ij}$ denotes the magnetic field strength of the $i$th row and the $j$th column of a synoptic map $B$.
Now, let $\mathcal{M}(B_0;t)$ denote the operator for SFT simulation.
It is a linear operator that takes in a synoptic map and evolves it to time $t$, so that the magnetic field at time $t$ is $B(t)=\mathcal{M}(B_0;t)$, where $B_0$ is the initial synoptic map.
Now, due to the linearity, we have 
\begin{equation}\label{eq:MatrixSim}
B(t)=\mathcal{M}(B_0;t)=\mathcal{M}(\sum_{ij} B_0^{ij}E_{ij};t) = \sum_{ij} B_0^{ij}\mathcal{M}(E_{ij};t).
\end{equation}
This equation shows that instead of simulating the evolution of the magnetic field with the SFT model, we can simulate the evolution of a set of matrix units $E_{ij}$, scale these simulations with the initial magnetic field values $B_0^{ij}$, and finally sum these together to produce the time evolution for the full synoptic map.
The advantage is that the matrix units need to be simulated only once, and the resulting matrices can then be used to simulate the evolution of any synoptic map with the time resolution that was used to calculate $\mathcal{M}(E_{ij};t)$.

If, as often is, the synoptic map has $180 \times 360 = 64800$ pixels, a general situation would require as many SFT simulations to produce the required $\mathcal{M}(E_{ij};t)$ matrices.
However, the equations of the SFT model described in Appendix~\ref{appendix:SFT} have considerable redundancy because of the large-scale flows that are both axisymmetric and antisymmetric across the equator.
For this reason, we only need to use the SFT model to simulate $\mathcal{M}(E_{ij};t)$ matrices for a single longitude of one hemisphere, which reduces the number of required SFT simulations down to 90.
All the remaining $\mathcal{M}(E_{ij};t)$ matrices can then be obtained via longitudinal translations and/or mirroring across the equator from this core set of 90.
Together the full set of $\mathcal{M}(E_{ij};t)$ matrices corresponds to the matrix $\mathbf{M}(t)$ in Eqs.~\ref{eq:SFTProduct} and ~\ref{eq:VSProduct}.

Let $\mathbf{B_0}$ and \textbf{B} denote the vectorized versions of the initial and final synoptic maps and \textbf{M}($t$) a matrix whose columns are equal to vectorized versions of $\mathcal{M}(E_{ij};t)$ matrices.
Equation \ref{eq:MatrixSim} can then be written as matrix multiplication shown in Eq.~\ref{eq:SFTProduct}.
The matrix \textbf{M}($t$) depends on time step used to create matrix unit responses $\mathcal{M}(E_{ij};t)$ .

The feasibility of the propagator approach depends on the size of matrix $\textbf{M}$.
It is a square matrix whose linear size equals the number of pixels in a synoptic map.
For a typical $180\times360$ synoptic map the size of $\textbf{M}$ is $64800\times64800$, which takes more than 30~Gb to store in memory.
In practice it is not necessarily efficient, or even possible, to calculate the matrix product in single step and the calculation needs to be carried in parts.
For our laptop setup, the calculation of this matrix product takes about the time of SFT simulation lasting two Carrington rotations.
Next we show, how the vector sum can be used to significantly reduce the dimensions of $\textbf{M}$, producing a matrix that represents the evolution of the solar dipole evolving under the SFT model.

\subsection{Propagator for the dipole flux transport}
Let $\mathcal{V}(B)$ denote the operator for the vector sum operation described in Appendix~\ref{appendix:VectorSum}.
It is a linear operator that takes a synoptic map $B$ and produces a three-component vector describing the magnitude and orientation of the solar dipole.
Now, due to the linearity of $\mathcal{V}$ and $\mathcal{M}$ the dipole vector at time $t$ can be calculated as
\begin{gather}
\begin{aligned}\label{Eq:Dipole}
V &= \mathcal{V}(B(t))=\mathcal{V}(\mathcal{M}(B_0;t)) = \mathcal{V}(\sum_{ij} B_0^{ij}\mathcal{M}(E_{ij};t)) \\
    &= \sum_{ij} B_0^{ij}\mathcal{V}(\mathcal{M}(E_{ij};t)),
\end{aligned}
\end{gather}
which corresponds to the matrix product of Eq.~\ref{eq:VSProduct}.
Equation \ref{Eq:Dipole} shows that to produce the evolution of the dipole vector, we do not need the full matrix \textbf{M}, but it suffices to work with more compact propagation matrix \textbf{VM}.
We can obtain the matrix \textbf{VM}  directly by calculating the vector sums of $\mathcal{M}(E_{ij};t)$ matrix unit responses.
For $180\times360$ synoptic map, the size of \textbf{VM} is $3\times64800$, which decreases the number of stored components by a factor of 21600.

Since multiplication by matrix \textbf{VM}(t) maps the synoptic map to dipole vector at single time $t$, it is practical to produce a set of matrices \textbf{VM}(t) such that $t\in \{t_1,...,t_n\}$.
This is straightforward to achieve as matrices corresponding to different timesteps can be created during the same SFT simulation.
We have computed the \textbf{VM}(t) matrices with daily and Carrington resolution up to 11 years.
Since the propagator matrices \textbf{VM}(t) describe the evolution of solar dipole under the dynamics of the SFT model we refer to method of simulating the solar dipole with Eq.~\ref{eq:VSProduct} as the dipole flux transport (DFT).

\section{Applications}\label{sec:Examples}
Here we compare the performance of the DFT and SFT simulations in a few different scenarios.
All simulations were performed on a Lenovo ThinkPad P52 laptop from 2019 equipped with an Intel Core i7-8750H CPU, 32 GB RAM, running Windows 11.
The computations were carried out using MATLAB R2022b without parallelization.
We measured the performance using MATLAB’s \texttt{timeit} function, which essentially reports the median runtime of 10 simulation runs.
Table~\ref{Table:DipoleSimulations} gives runtimes for the SFT and DFT simulations for different scenarios.
In the last two rows, the SFT runtimes are estimated from the solar cycle 24 SFT simulation, as running them in full would require excessive computation time.
The propagation matrices and code for running the following tests are available on GitHub\footnote{\url{https://github.com/itahtine/DipoleFluxTransport}}.

\begin{table*}
\caption{Runtimes of different simulation scenarios.}             
\centering            
\begin{threeparttable}
\begin{tabular}{c c c c c}
\hline\hline                   
Scenario & Resolution & SFT Runtime (s) & DFT Runtime (s) & Speedup factor \\   
\hline              
    Single map, 4 years & CR & 259.41 & 0.09 & 2982 \\     
    Single map, 4 years & Daily   & 259.41 & 2.18 & 119 \\  
    Solar cycle 24 & Daily         & 608.16 & 7.48 & 81 \\
    Solar cycle 24, individual ARs & Daily & 411030.08\tnote{*}  & 7.48 & 54982  \\
    Ensemble prediction, 180 days   & Daily & 4103.47\tnote{*}  & 0.74 & 5392  \\ 
\hline                  
\end{tabular}
\label{Table:DipoleSimulations}
\begin{tablenotes}
\item[*] Estimated based on the runtime of the SFT simulation of solar cycle 24
\end{tablenotes}
\end{threeparttable}
\end{table*}

\subsection{Evolution of active regions}
The simplest and most straightforward application is the evolution of single map without any flux emergence.
Such simulations are useful, for example, for studying how the properties of the active regions affect their evolution. 
The runtime of the DFT simulation depends on the number of non-zero pixels as we only need to compute their time evolution.
Because the runtime of the DFT depends on number of non-zero pixels in initial map, we selected the largest active region of solar cycle 24 corresponding to NOAA 12192 to obtain the estimate for the upper bound of the runtime of the DFT simulation. 
We ran the simulation for four years, which is enough time for the nonaxisymmetric magnetic field to annihilate and magnetic field to settle into an asymptotic axial dipole state \citep[see, e.g.,][]{WangSheeley2002}.

With Carrington resolution a 4-year (54 rotation) DFT simulation of an active region takes 0.09 seconds.
DFT speeds up the calculation by a factor of 2982 when compared to SFT simulation of the same length that takes 259.41 seconds.
Increasing time resolution slows down the matrix calculation roughly linearly, and with the daily resolution the similar 4-year simulation takes 2.18 seconds.
Even with daily resolution the DFT performs 119 times faster than the SFT.

\subsection{Evolution of solar cycle}\label{sec:MultipleRegionsExample}
DFT can also be applied to multiple active regions.
To do so the evolution of each active region and the initial magnetic field needs to be simulated independently.
Each simulation produces a vector timeseries that, when summed together, produce the evolution of the full dipole vector of the solar magnetic field.
For this summation to make sense, the time resolution needs to be sufficiently high so that the individual timeseries, which have different initial starting points, can be lined up together.
For global simulations of the solar magnetic field which typically have a time resolution of one Carrington resolution, a 1 day resolution for individual simulations seems a reasonable choice.
Since the evolution of each active region needs to be simulated separately, the cost of such simulations grows with the number of active regions.

\begin{figure}[!htbp]
\resizebox{\hsize}{!}{\includegraphics{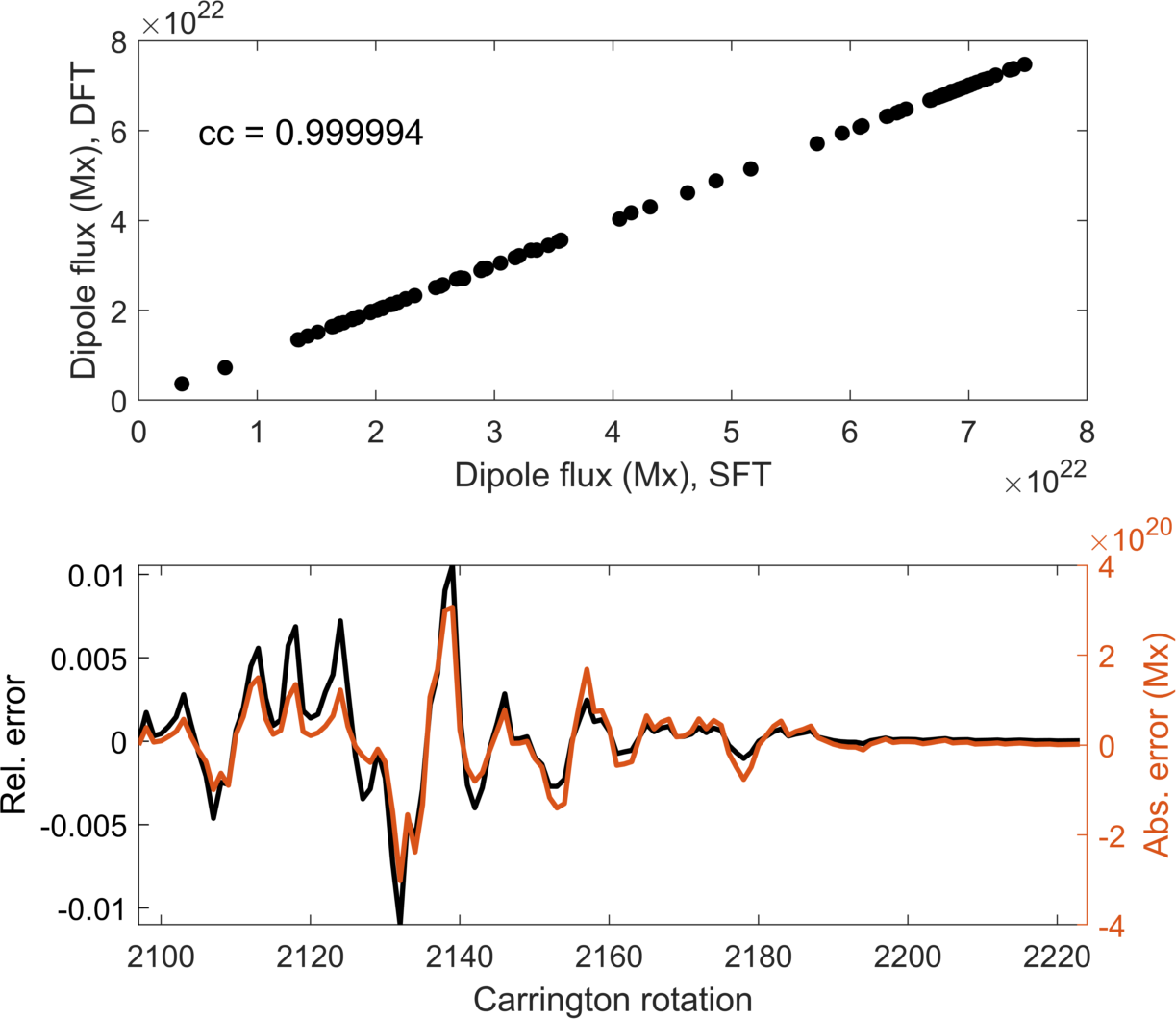}}
\caption{Comparison of SFT and DFT simulations. Upper panel: Scatter plot showing the dipole vector magnitude from the DFT (y-axis) and from the SFT (x-axis) simulation. Lower panel: Relative (black) and absolute (right) error of the DFT dipole vector magnitude.
}\label{fig:DipoleSim}
\end{figure}

Upper panel of Fig.~\ref{fig:DipoleSim} compares the dipole vector magnitude between the DFT and the SFT simulations for Carrington rotations 2097-2224.
The Pearson correlation between the two methods is 0.999994.
Lower panel shows that the relative difference between the DFT and the SFT simulation is at most 1\%.
The DFT simulation takes only 7.48 seconds in our tests which outperforms SFT simulation lasting 608.16 seconds by factor of 81.

For some applications it is useful to have the time evolution of individual active regions \citep[e.g.,][]{Yeates2020,Tahtinen2026}, which the DFT simulation gives for free.
In this case we estimate that the DFT performs 54982 times faster than the SFT model which would take about 4.76 days because the evolution of each of the 1095 active regions needs to be simulated separately.

\subsection{Ensemble predictions}
Suppose that we are interested in creating an ensemble forecast of the solar magnetic field using active regions generated from some probabilistic model for some time ahead and we want to study the performance of our model for solar cycle 24.
We can do this by generating an ensemble of $N$ realizations of active regions and simulating the magnetic field by $t$ amount of time  forward starting from each Carrington rotation between CR2097-CR2224 (128 rotations in total).
Assuming that the number and timing of active regions on average matches with the observed active regions, we can use the HMI SHARPS to estimate the runtime of such prediction.
The median runtime for such a 180-day hindcast with the DFT is 0.74 seconds while for the SFT simulation the estimated time is 4103 seconds.
Taking $N = 100$, the DFT runs in about a one minute while the SFT simulation takes over 4.5 days.

\section{Discussion}\label{sec:Discussion}
The DFT model presented in this Letter allows us to simulate the evolution of the solar dipole extremely efficiently.
This is made possible for two reasons.
First, linearity of the classic SFT model allows us to precompute the propagation matrices that capture the dynamics of the SFT model allowing us to turn the SFT simulation into matrix product.
Second, the significant speedup of computations is achieved by using the vector sum to compress the size of propagation matrices by a factor of 21600.

For a single map (e.g., active region), the DFT provides numerically identical results with the SFT model at the time resolution of the propagation matrix.
The DFT can also be used to model the evolution of the solar dipole subject to multiple source regions emerging at different times.
In our tests, daily resolution propagation matrix is enough to produce results that agree with Carrington resolution SFT simulation with maximum relative error around 1\%.
Compared to earlier methods used to speed up the simulations of the solar dipole \citep{DeVore1984,Cameron2007,Ijima2017,Petrovay2019,Yeates2020,Petrovay2020,Wang2021,Tahtinen2026}, the DFT produces the evolution of the full dipole vector, not just its axial component, and it works for arbitrary initial conditions.

DFT shares the same limitations as the SFT model from which it was derived and cannot account for nonlinear processes, such as reduced effective diffusivity in regions of strong magnetic field. In principle, DFT can be applied to time-dependent scenarios (e.g., varying meridional flow amplitude), but this can be rather cumbersome because the propagator matrices become path-dependent.

Despite its simplicity, the solar dipole contains a lot of crucial information about the large-scale magnetic field, making the efficient simulations thus highly useful.
The speedup is greatest in situations that demand many individual simulations.
For example, \citet{Tahtinen2026} simulated the evolution of individual active regions to quantify their effect on the large-scale dipole.
With the DFT, this calculation is 1000 times faster than their calculation which was done parallel on a computer cluster with 40 cores.
Another context in which the DFT is likely to prove useful is the ensemble modeling.
For example, given a generative active region model \citep[see, e.g.,][]{Jha2024}, the DFT can be used for probabilistic ensemble modeling extremely efficiently.
In addition, as the magnitude of the dipole vector closely matches the PFSS OSF \citep{Tahtinen2024}, the development of OSF can be studied in various scenarios with almost negligible computational cost.

\section{Conclusions}\label{sec:Conclusions}
We have developed a DFT matrix method to simulate the evolution of the solar dipole up to 50000 times faster than regular SFT simulations.
Importantly, the method provides the full dipole vector, greatly increasing the scope of dipole simulations, which have so far been limited to the axial dipole component.
We anticipate that the DFT will prove especially useful for ensemble methods to study solar activity.
It can also be used to study the development of OSF extremely efficiently in various scenarios.

\begin{acknowledgements}
I.T. and T.A. acknowledge the financial support by the Research Council of Finland to the SOLEMIP (project no. 357249).
We thank the anonymous referee for their constructive comments.
The authors wish to acknowledge CSC – IT Center for Science, Finland, for computational resources.
\end{acknowledgements}

\bibliography{bibliography}

\begin{appendix}
\section{Surface flux transport}\label{appendix:SFT}
The classic SFT model \citep[see, e.g.,][]{Yeates2023} describes the evolution of the radial magnetic field subject to differential rotation $\Omega(\theta)$, meridional flow $u_\theta(\theta)$, and turbulent diffusion $\eta$ with the induction equation
\begin{equation}
\frac{\partial B_r}{\partial t} + \nabla_h \cdot (\mathbf{u}_h B_r) = \eta \nabla_h^2 B_r + S,
\end{equation}
where $\mathbf{u}_h$ describes the horizontal flow velocity (due to $\Omega(\theta)$ and $u_\theta(\theta)$) and the magnetic flux emergence is modeled with the source term S.
The explicit form of the governing equation is
\begin{gather}
\begin{aligned}
    \frac{\partial B_r}{\partial t} = &-\frac{1}{R_{\odot} \sin \theta} \frac{\partial}{\partial \theta} \left( \sin \theta \, u_{\theta} B_r \right)
    - \Omega(\theta) \frac{\partial B_r}{\partial \phi} \\
    &+ \frac{\eta}{R_{\odot}^2 \sin \theta} \frac{\partial}{\partial \theta}  
    \left( \sin \theta \frac{\partial B_r}{\partial \theta} \right)
    + \frac{\eta}{R_{\odot}^2 \sin^2 \theta} \frac{\partial^2 B_r}{\partial \phi^2} + S.
\end{aligned}
\end{gather}

In this Letter, we adopt the differential rotation profile of \citet{Snodgrass1990}
\begin{equation}
    \Omega(\theta) = 0.18 - 2.396 \cos^2(\theta) - 1.787 \cos^4(\theta) \quad [^{\circ} \text{day}^{-1}],
\end{equation}
and the meridional flow profile of \citet{Whitbread2018}
\begin{equation}
        u_{\theta}(\theta) = -u_0 \sin^p{\theta}\cos{\theta},
\end{equation}
with their shape parameter $p = 2.33$.
We set diffusivity to $\eta=350~\mathrm{km^2/s}$ and the meridional flow amplitude to $u_0=11$~m/s, for which \cite{Tahtinen2026} found the best-fit values for solar cycle 24, using the total vector sum as optimization metric.

\section{Vector sum and solar dipole}\label{appendix:VectorSum}
The solar dipole can be efficiently represented using the recent vector sum method \citep{Tahtinen2024,Tahtinen2026}.
In the vector sum method, the pixels of a synoptic magnetogram are represented as vectors in heliographic spherical coordinates.
The length of each vector corresponds to the total magnetic flux within the magnetogram pixel and the direction to the location of the pixel on the solar surface.
The vector sum is then straightforwardly a sum of these spherical vectors producing a single vector describing the orientation and strength of the solar dipole.
The orientation of the dipole vector equals the orientation of the first multipole (dipole) in the spherical harmonic expansion, but its magnitude differs from the solar dipole moment by a factor of $\frac{4\pi R_\odot^2}{3}$ and it has units of flux instead of flux density.

The magnitude of the dipole vector closely matches the open solar flux (OSF) derived from the potential field source surface (PFSS) model with a source surface radius $R_{ss}=2.5R_\odot$.
Furthermore, \citet{Tahtinen2026} show that the magnitude of the dipole vector equals the photospheric magnetic flux aligned with the dipole axis, thus relating PFSS OSF directly to the distribution of the magnetic fields in the photosphere.
\end{appendix}

\end{document}